\documentclass[journal]{IEEEtran}

\usepackage{myStyleIEEE}
\usepackage{ amssymb }
\usepackage{mathtools}
\usepackage{graphicx}

\IEEEoverridecommandlockouts

\begin{document}

\title{Enhanced Modelling Framework for Equivalent Circuit-Based Power System State Estimation  \vspace*{-2mm}}

\renewcommand{\theenumi}{\alph{enumi}}
\renewcommand{\arraystretch}{1.3}
\author{
Aleksandar Jovicic, \textit{Student Member, IEEE}, Marko Jereminov, \textit{Student Member, IEEE},\\ Larry Pileggi, \textit{Fellow, IEEE}, Gabriela Hug, \textit{Senior Member, IEEE}
\vspace*{-0.525cm}}

\maketitle
\IEEEpeerreviewmaketitle

            
\begin{abstract}
The equivalent split-circuit formulation is a novel approach that has recently been applied to a range of power system related problems. As a result, a linear and a nonlinear method for power system state estimation with simultaneous treatment of the conventional and synchrophasor measurements have been proposed. In this paper, new circuit models are introduced for different combinations of conventional measurements, thus providing a complete modelling framework for these methods. Additionally, handling cases of null injections and buses with no measurements are included. Simulations are performed on several test systems in order to evaluate the performance of both methods with the enhanced modelling framework and to compare them to a hybrid constrained estimator based on the conventional WLS approach.\\
\end{abstract}

\begin{IEEEkeywords}
State estimation, equivalent circuit formulation, phasor measurement units, conventional measurements, power system modelling
\end{IEEEkeywords}

\section{Introduction}
The operating state of a power system has to be monitored continuously in order for a system operator to have an insight into the actual condition of the grid at all times. For this purpose, various measurement devices are installed throughout the grid to observe different physical quantities, and the measured values are then transmitted to a supervisory center via communication links. However, the so obtained measurement data contain errors due to imperfections of the measurement equipment, noise and delays in the communication channels, etc. Consequently, State Estimation (SE) is utilized to provide an estimate of the most probable state of the system, which is commonly expressed as a set of voltage magnitudes and angles for all buses in the grid \cite{Abur}.    

For a long time, power systems have been monitored by measurements of bus voltage magnitudes, active and reactive power injections at the buses, as well as active and reactive line flows, which are processed by Remote Terminal Units (RTU). The Weighted Least Square (WLS) algorithm that was formulated in \cite{Schweppe} is one of the most frequently used methods for state estimation based on these types of data. It is founded on minimizing the square of the difference between each measurement and its corresponding measurement function that relates the measurement to the system states. Since measurement functions are derived from power flow equations, which are nonlinear, an iterative approach must be conducted in order to solve the problem. 

Recent advances in the area of synchronized measurements and increased utilization of Phasor Measurement Units (PMU) in transmission networks has significantly changed the available measurement set. Very high measurement accuracy of these devices and their synchronization by the Global Positioning System (GPS) allow for an improved performance of state estimators. The set of PMU measurements comprises voltage and current phasors. As a result, if a sufficient number of PMUs is installed in the grid to provide full system observability by the phasor measurements, a linear SE algorithm can be utilized to provide a highly accurate estimate of the system state \cite{Phadke1985}. The main obstacle for increased penetration of PMUs in the present day transmission systems is the cost associated with their installation \cite{DOE}. Thus, conventional RTU measurements are still needed in order to preserve full system observability. Therefore, it is required to derive a hybrid state estimation (HSE) method that will account for both synchrophasor and conventional measurements, thus taking advantage of the high precision of the former, and substantial grid coverage by the latter. One of the main challenges is the integration of current phasor measurements, which can jeopardize convergence properties and worsen numerical conditioning of the estimation problem \cite{Valverde2011}.

One distinguished group of HSE algorithms are multi-stage estimators that are characterized by a sequential treatment of RTU and PMU measurements \cite{Ming2006,Baltensperger2010,Nuqui}. Division of different types of measurements allows for separate SE formulations at each stage, depending on the type of measured quantities that are processed. An alternative, yet similar approach is proposed in \cite{Costa2013}, where PMU and RTU measurements are treated by separate estimators that are executed in parallel. Their estimates are then combined by a fusion technique to obtain the final estimate of the system state. Another group of algorithms is HSEs that combine both types of measurements into a single-stage formulation. Methods in \cite{Bi2008,Chakrabarti2010,Asprou2011} include some form of transformation of the current phasor measurements, while the algorithm derived in \cite{Valverde2011} utilizes all measurements directly, which is enabled by expanding the vector of states to include current magnitudes and angles. All techniques proposed in \cite{Ming2006,Baltensperger2010,Nuqui,Costa2013,Bi2008,Chakrabarti2010,Asprou2011,Valverde2011} employ WLS-based estimators and are therefore solved iteratively due to their inherent nonlinearity. Apart from WLS-based methods, some of the most recent alternative approaches include HSE based on the Least Absolute Value method \cite{Moreno2018} and decentralized HSE \cite{Gol2018}. In all of these cases the inherent nonlinearity of the measurement functions causes a significant computational burden. 

To address the challenges imposed by the nonlinearity of the equations already present in the power flow problem, efforts have been made to employ concepts of circuit theory to a range of power system problems, which resulted in the Equivalent Circuit Formulation (ECF) of the power flow problem \cite{CMU2,CMU3,CMU4,Pandey}. The main idea of this approach is that any power system component can be represented in terms of its relation between voltage and current at its terminals, and subsequently mapped into a circuit that accurately depicts this relation. This modelling framework can be used in combination with various circuit simulation tools \cite{Pileggi} to address a range of power system related problems.

The fact that ECF is based on voltage and current state variables in rectangular coordinates lends itself to the derivation of novel SE algorithms, and is particularly suitable for the inclusion of PMU measurements as will be shown later. A nonlinear ECF-based estimator (NECF-SE) was proposed in \cite{Jovicic}. Circuit models for PMU measurements (linear model) and RTU measurements of voltage magnitude and active and reactive injected powers (nonlinear model) were derived to enable circuit representations of the sensed data and integrate them into the circuit representation of the grid. Subsequently, an optimization problem was formulated to obtain the estimate of the system state in rectangular coordinates. 

Based on the NECF-SE method, a novel linear ECF-based estimator (LECF-SE) was then developed and presented in \cite{Jovicic2}. Instead of the nonlinear RTU model that was used in NECF-SE, a linear current source-based model was introduced to allow for a fully linear modelling framework for the entire system, taking advantage of the fact that the circuit models of transmission lines, transformers, phase-shifters and shunts that have been derived previously in \cite{CMU2,CMU3,CMU4,Jovicic} are linear as well. This allowed for the states to be estimated by solving an optimization problem with linear first-order optimality conditions, thus significantly reducing the computational burden of this formulation and making it suitable for real-time SE.   

 The PMU circuit models that were used in \cite{Jovicic} and \cite{Jovicic2} can represent any configuration of phasor measurements observed at a bus. However, the RTU circuit models that were derived can only capture the relation between bus voltage magnitude and injected active and reactive power, i.e. line flow measurements could not be included explicitly, thereby neglecting available information that could significantly improve the estimation accuracy. In this paper, the circuit modelling frameworks of ECF-based estimators are further enhanced to include circuit models for different combinations of RTU measurements, including bus injection and line flow measurements, and the possibility of having no measurements at a bus. As a result, a complete library of RTU models is created with a suitable circuit model for each combination of measurements of voltage magnitude, active and reactive power injection, and active and reactive power flow. Moreover, this paper compares the overall performance of both NECF-SE and LECF-SE algorithms including these complete models by implementing them on several test cases. Additionally, since both NECF-SE and LECF-SE are essentially constrained estimators, their performance is also compared to a well-known constrained HSE algorithm, proposed in \cite{Valverde2011}, which is a conventional WLS-based method.

The remainder of the paper is structured as follows. Section \ref{sec: 3} provides an overview of the equivalent split-circuit formulation of the power flow problem, and summarizes the previous work in the area of state estimation based on ECF. In Sect. \ref{sec: 4}, new circuit models are introduced, thus providing a complete modelling framework for both ECF-based estimators. Section \ref{sec: 5} discusses the simulation results, and the paper is concluded in Sect. \ref{sec: 6}.    
\section{ECF-Based State Estimation} \label{sec: 3}
\subsection{Equivalent Circuit Formulation}
The equivalent circuit formulation, proposed in \cite{CMU2,CMU3,CMU4}, is a novel approach to model power systems. Instead of relying on the standard power flow equations, the main idea of ECF is that an entire power system can be modelled in terms of voltage and current state variables in rectangular coordinates and represented as an electric circuit. This is achieved by mapping each power system component to an equivalent circuit model, which accurately depicts the relation between voltage and current at its terminals. Interconnection of the so obtained circuit models yields an equivalent circuit of the entire system. Governing equations of this circuit can be formulated by applying some of the well-known circuit-based formulations, such as Modified Nodal Analysis \cite{Pileggi}. Ultimately, the operating point of the circuit that is found by solving the circuit equations represents the power flow solution.

Since many power system components are characterized by nonlinear and non-differentiable equations, the relations between voltages and currents at the terminals of each power system component have to be split into real and imaginary parts to provide analyticity. This is equivalent to splitting the equivalent circuit of the system into two coupled sub-circuits, where the first sub-circuit comprises real voltages and currents, and the second represents their imaginary parts \cite{CMU2}. By doing so, the governing equations of the system's circuit can be linearized and iteratively solved by applying the Newton-Raphson method. 

\begin{figure}[t!]
\vspace{-0.7cm}
\centering
\begin{subfigure}{0.45\textwidth}
\includegraphics[width=\textwidth]{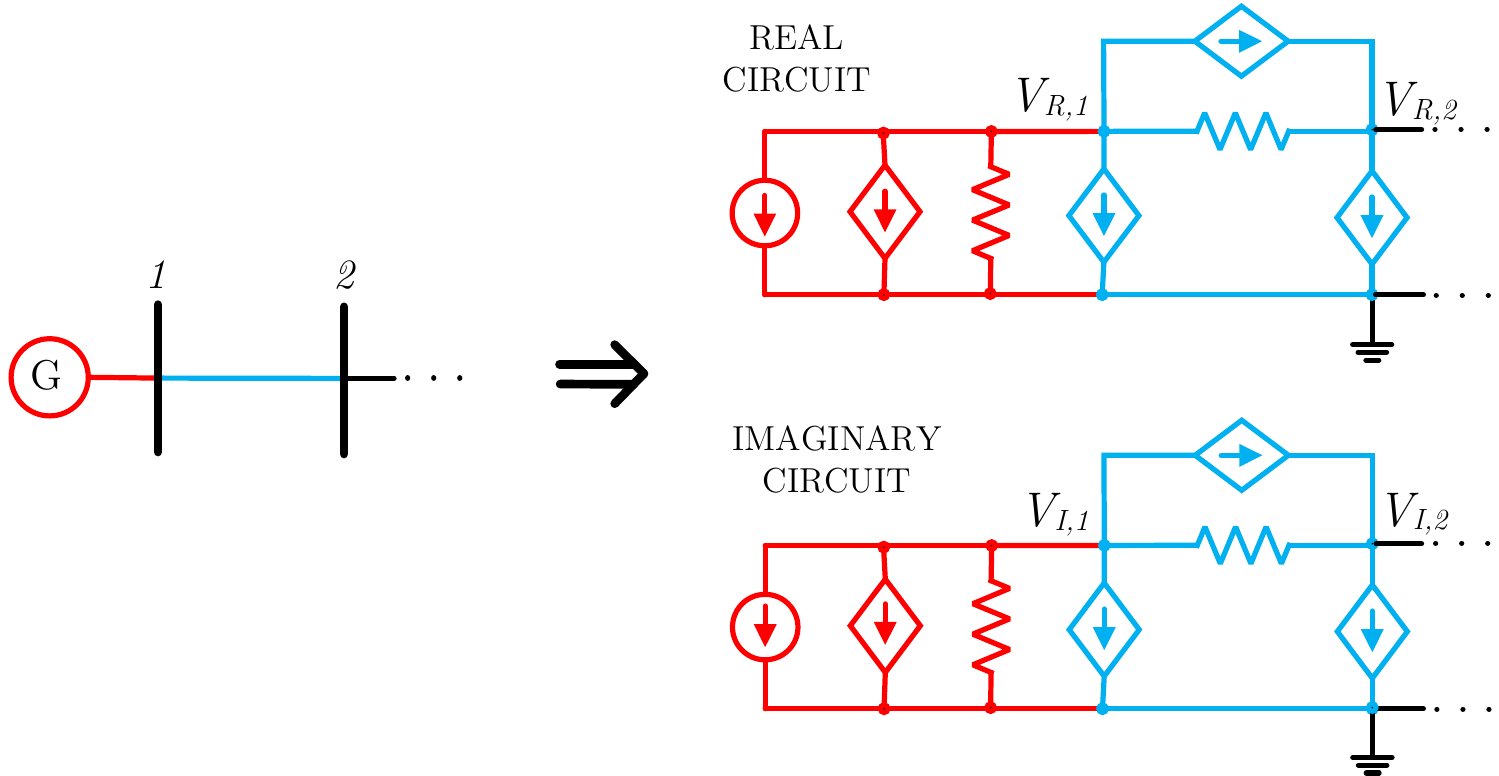}
\end{subfigure}
\caption{Equivalent split-circuit of a generator connected to a transmission line \cite{CMU2}.}
\label{fig:2bus}
\vspace{-0.3cm}
\end{figure}

As an example, the linearized equivalent circuit for a small fraction of a system consisting of a generator connected to a transmission line is shown in Fig. \ref{fig:2bus}.
The corresponding system components and their equivalent sub-circuits are colour-coded. The current of each controlled current source in the real sub-circuit is a function of the voltage across its counterpart in the imaginary sub-circuit, and vice versa. The generator circuit model (red) is obtained by using linearization. On the contrary, a transmission line (blue) is directly mapped to the linear circuit model. This is clearly an advantage of this modelling approach over the inherently nonlinear modelling framework used for the transmission network if the power mismatch equations are used. More details on the derivation of the sub-circuits shown in Fig. \ref{fig:2bus} can be found in \cite{CMU2}. 

\subsection{ECF-Based State Estimators} \label{subsec: 3a}
ECF has been leveraged recently to formulate a novel approach for SE in transmission systems, hence leading to the introduction of NECF-SE \cite{Jovicic,Jereminov} and LECF-SE \cite{Jovicic2} estimators. A brief overview of their similarities and differences will be presented below. 

\begin{figure}[b!]
    \vspace{-0.175cm}
    \centering
    \begin{subfigure}[h]{0.135\textwidth}
        \centering  
        \includegraphics[width=\textwidth]{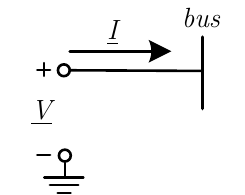}
        \caption{\hspace*{0.25em}}  
        \label{fig:PMU_Inj_a}
    \end{subfigure}
    \hfill
    \begin{subfigure}[h]{0.165\textwidth}  
        \centering 
        \includegraphics[width=\textwidth]{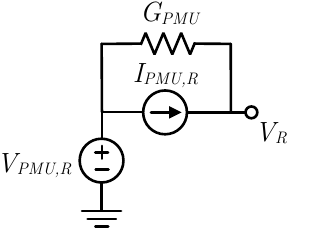}
        \caption{\hspace*{0.25em}}  
        \label{fig:PMU_Inj_b}
    \end{subfigure}
    \hfill
        \begin{subfigure}[h]{0.165\textwidth}  
        \centering 
        \includegraphics[width=\textwidth]{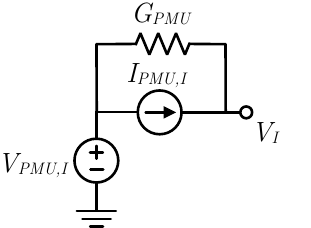}
        \caption{\hspace*{0.25em}}  
        \label{fig:PMU_Inj_c}
    \end{subfigure}
    \caption{PMU injection model: (a) measurement data; (b) real circuit; (c) imaginary circuit.}
    \label{fig:PMU_inj}
\end{figure}

In ECF-based estimators all measurements are translated into a suitable circuit model. As an example, the circuit model used for a bus that is equipped with a PMU measuring voltage and injected current phasor is given in Fig. \ref{fig:PMU_inj}.  
Voltage and current measurements are represented by independent voltage and current sources, respectively. Conductance $G_{PMU}$ is added in order to preserve the effect of both voltage and current source on the rest of the system's equivalent circuit. Also, $G_{PMU}$ serves as a slack element, since a non-zero current flowing through it indicates deviation of the measured data from the actual state of the system. Consequently, NECF-SE and LECF-SE algorithms are based on minimization of the currents flowing through the PMU conductances. The PMU model shown in Fig. \ref{fig:PMU_inj} is used in both NECF-SE and LECF-SE formulations. The same modelling approach is also used in case PMU measurements of branch currents are available. Then, each branch current is represented as a parallel connection of a current source and a conductance. Therefore, all possible sets of PMU data are covered by the appropriate circuit models that were already introduced in \cite{Jovicic,Jovicic2}.

RTU circuit models were derived under the assumption that the set of RTU measurements comprises voltage magnitude, current magnitude and power factor data, since these are originally measured quantities and hence contain smaller errors \cite{Caro}. In practice, values of active and reactive power are usually communicated instead of current magnitude and power factor measurements, but these types of data can still be mapped to the same circuit models, based on the following relations between bus voltages and injection currents in rectangular coordinates:
\begin{align}
    I_{R} &= \frac{I}{V}\text{cos}\left(\phi\right)V_{R} + \frac{I}{V}\text{sin}\left(\phi\right)V_{I} = \frac{P}{V^2}V_{R} + \frac{Q}{V^2}V_{I} \label{eq: I_R}\\
    I_{I} &= \frac{I}{V}\text{cos}\left(\phi\right)V_{I} - \frac{I}{V}\text{sin}\left(\phi\right)V_{R} = \frac{P}{V^2}V_{I} - \frac{Q}{V^2}V_{R} \label{eq: I_I}
\end{align}
where $V_R$ and $V_I$ are real and imaginary bus voltages, $I_R$ and $I_I$ are real and imaginary injection currents, $V$ is the measured voltage magnitude, $I$ the measured injection current magnitude, cos$(\phi)$ the measured power factor, and $P$ and $Q$ are measured active and reactive power injections, respectively. These equations hold under the assumption that the reference directions of the measured voltage and injected apparent power correspond to load conditions. The reader is referred to \cite{Jovicic} for more details on the derivation of \eqref{eq: I_R}-\eqref{eq: I_I}. 

\begin{figure}[t!]
    \vspace{-0.175cm}
    \centering
    \begin{subfigure}[h]{0.135\textwidth}
        \centering  
        \includegraphics[width=\textwidth]{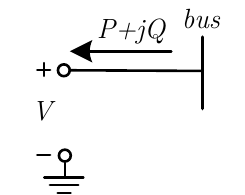}
        \caption{\hspace*{0.25em}}  
        
    \end{subfigure}
    \hfill
    \begin{subfigure}[h]{0.165\textwidth}  
        \centering 
        \includegraphics[width=\textwidth]{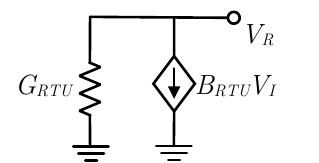}
        \caption{\hspace*{0.25em}}  
        \label{fig:RTU_interval_real}
    \end{subfigure}
    \hfill
        \begin{subfigure}[h]{0.165\textwidth}  
        \centering 
        \includegraphics[width=\textwidth]{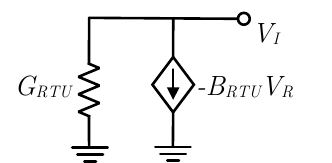}
        \caption{\hspace*{0.25em}}  
        \label{fig:RTU_interval_imag}
    \end{subfigure}
    \caption{Nonlinear RTU injection model: (a) measurement data; (b) real circuit; (c) imaginary circuit.}
    \label{fig:RTU_interval_inj}
\end{figure}

\begin{figure}[t!]
    \vspace{-0.175cm}
    \centering
    \begin{subfigure}[h]{0.135\textwidth}
        \centering  
        \includegraphics[width=\textwidth]{images/RTUdata.pdf}
        \caption{\hspace*{0.25em}}  
        \label{fig:RTU}
    \end{subfigure}
    \hfill
    \begin{subfigure}[h]{0.165\textwidth}  
        \centering 
        \includegraphics[width=\textwidth]{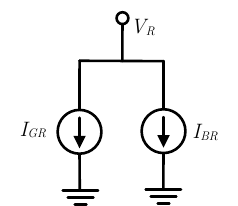}
        \caption{\hspace*{0.25em}}  
        \label{fig:RTU_Real}
    \end{subfigure}
    \hfill
        \begin{subfigure}[h]{0.165\textwidth}  
        \centering 
        \includegraphics[width=\textwidth]{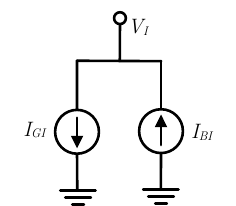}
        \caption{\hspace*{0.25em}}  
        \label{fig:RTU_Imag}
    \end{subfigure}
    \caption{Linear RTU injection model: (a) measurement data; (b) real circuit; (c) imaginary circuit.}
    \label{fig:RTU_lin_model}
\vspace{-1em}
\end{figure}

One of the key differences between the two ECF-based estimators is the way relations \eqref{eq: I_R}-\eqref{eq: I_I} are translated to a circuit model for RTU measurements. In NECF-SE, the set of RTU measurements is mapped to the admittance-based circuit model that is shown in Fig. \ref{fig:RTU_interval_inj}. Since all node voltages, as well as $G_{RTU}$ and $B_{RTU}$ components, are variables in the state estimation problem, the given circuit model is nonlinear. More details on the derivation and utilization of this model can be found in \cite{Jovicic}. In order to reduce mathematical complexity imposed by the nonlinear modelling of RTU measurements, the LECF-SE algorithm translates equations \eqref{eq: I_R}-\eqref{eq: I_I} into the linear circuit model comprising independent current sources, given in Fig. \ref{fig:RTU_lin_model}. The variables in this model are the values of the current sources. More details on the derivation of this circuit model can be found in \cite{Jovicic2}.

An important observation is that both models from Fig. \ref{fig:RTU_interval_inj}-\ref{fig:RTU_lin_model} can be directly applied only in two cases. The first is if the measurements of injected active and reactive power are the only available power measurements at an RTU bus, and the second is if the set of available RTU power data consists of line flow measurements in all lines adjacent to the observed RTU bus, which allows transforming this information into pseudo-measurements of the injected active and reactive power. In the latter case, significant information is lost as the information on individual line flows is aggregated into injections measurements. Providing models for cases with arbitrary combinations of measurements, i.e. partial injection and flow measurements, is therefore one of the key contributions of this paper.

The circuit models for the measurements are connected at the corresponding locations to the circuit that models the overall system including transmission lines, transformers, phase shifters and shunts as described in \cite{CMU2,CMU3,CMU4}. In NECF-SE and LECF-SE, the states are estimated by solving an optimization algorithm that includes the circuit equations as constraints and yields the most probable values of the circuit components representing the measurements. Voltages are estimated in rectangular coordinates, which is one of the important differences between traditional WLS- and ECF-based estimators. 

The NECF-SE algorithm estimates the system state by solving the following optimization problem:
\begin{subequations}
\label{eq: optimization}
\begin{align}
\min &\quad \lVert \boldsymbol{I_{G}}\rVert^2_2 + \lVert \boldsymbol{G_{diff}} \rVert^2_2 + \lVert \boldsymbol{B_{diff}} \rVert^2_2 \label{eq: obj_func}
\end{align}
\vspace{-0.5cm}
\begin{alignat}{6}
    &\text{s.t} \quad\quad\quad\quad && I_{ct}(\boldsymbol{X}) = 0 \label{eq: lin_eq}\\
    &&& \boldsymbol{X_{min}}\leq \boldsymbol{X}\leq \boldsymbol{X_{max}}\label{eq: var_Limit} 
\end{alignat}
\end{subequations}
where $\boldsymbol{X}$ is the vector of equivalent circuit state variables. The goal is to minimize the currents flowing through the PMU conductances, $\boldsymbol{I_G}$, and $\boldsymbol{G_{diff}}$ and $\boldsymbol{B_{diff}}$ that represent the difference between RTU sub-circuit variables and their corresponding measured values. The optimization problem is subject to a set of equality constraints \eqref{eq: lin_eq}, which represent the governing equations of the system's equivalent circuit, i.e. KCL equations for all nodes in the circuit, with some of them being nonlinear due to the modelling of RTUs. Aside from the different models for the RTU measurements, the NECF-SE differs from the LECF-SE by the fact that it treats measurement errors as intervals. As a result, its optimization problem comprises inequality constraints \eqref{eq: var_Limit} to bound the vector of state variables. More details about the optimization problem \eqref{eq: optimization} can be found in \cite{Jovicic}. Due to the structure of the problem, it was solved by using commercial nonlinear solvers. Alternatively, the problem can be solved by formulating its optimality conditions in terms of equivalent circuit constraints \cite{Jereminov}. In this way, the NECF-SE algorithm can be executed fully within the circuit domain.  

The LECF-SE algorithm on the other hand uses a weighted approach to account for measurement errors, hence resembling the conventional WLS method. The states are estimated by solving the following optimization problem: 
\begin{subequations}
\label{eq: optimization2}
\begin{align}
\min && \lVert \boldsymbol{I_{G}}\rVert^2_2 + \lVert \boldsymbol{V^{PMU}_{diff}} \rVert^2_2 + \lVert \boldsymbol{I^{PMU}_{diff}} \rVert^2_2 + \lVert \boldsymbol{I^{RTU}_{diff}} \rVert^2_2  \label{eq: obj_func2}
\end{align}
\vspace{-0.5cm}
\begin{equation}
    \text{s.t} \quad\quad\quad\quad  I_{ct}(\boldsymbol{X}) = 0 \label{eq: lin_eq2}\\
\end{equation}
\end{subequations}
where $\boldsymbol{X}$ is again the vector of equivalent circuit state variables. The terms that are minimized are the currents flowing through the PMU conductances, $\boldsymbol{I_G}$, the $\boldsymbol{V^{PMU}_{diff}}$ and $\boldsymbol{I^{PMU}_{diff}}$ terms that represent the weighted difference between PMU voltage and current variables and their corresponding measured values, as well as $\boldsymbol{I^{RTU}_{diff}}$ that represents the weighted difference between RTU current variables and their respective linear measurement functions. For more details, the reader is referred to \cite{Jovicic2}. Since the entire system along with the measurements is modelled by an equivalent linear split-circuit, the problem is subject to a set of linear equality constraints \eqref{eq: lin_eq2}, i.e. governing equations of the equivalent circuit, which further yields linear first-order optimality conditions for the given problem and thus substantially decreases its computational complexity.
\begin{figure*}[t!]
    \vspace{-0.5cm}
    \centering
    \begin{subfigure}[h]{0.2\textwidth}
        \centering  
        \includegraphics[width=\textwidth]{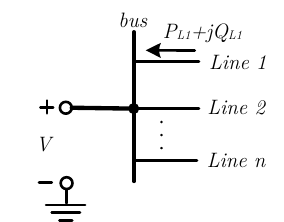}
        \caption{\hspace*{0.25em}}  
        \label{fig:1lineflow_meas}
    \end{subfigure}
    \hfill
    \begin{subfigure}[h]{0.34\textwidth}  
        \centering 
        \includegraphics[width=\textwidth]{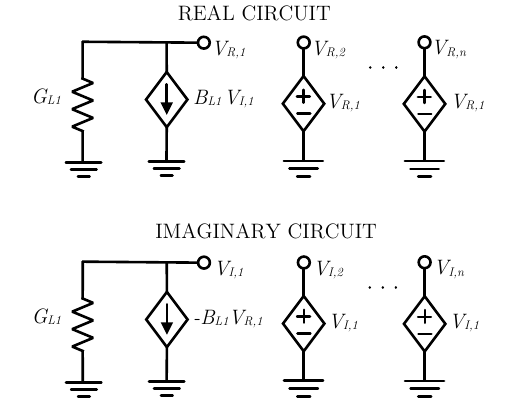}
        \caption{\hspace*{0.25em}}  
        \label{fig:1lineflow_int}
    \end{subfigure}
    \hfill
        \begin{subfigure}[h]{0.34\textwidth}  
        \centering 
        \includegraphics[width=\textwidth]{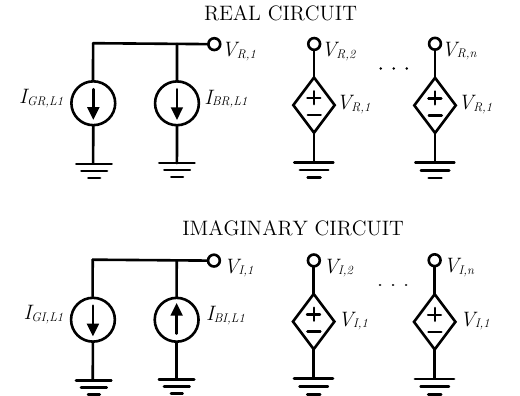}
        \caption{\hspace*{0.25em}}  
        \label{fig:1lineflow_lin}
    \end{subfigure}
    \caption{Circuit models for the case of line flow measurements in a single branch: (a) measurement data; (b) nonlinear circuit model; (c) linear circuit model.}
    \label{fig:1lineflow}
    \vspace{-1em}
\end{figure*}
\section{New Circuit Models for Complete Modelling of the Power Grid Measurements} \label{sec: 4}
As it was previously mentioned, aggregated circuit models incorporating the entire set of available measurements at a bus have been used in the ECF-based estimators previously. A modelling framework that would accurately incorporate all different sets of injection and line flow measurements potentially available at an RTU bus has not been fully developed yet. This is important as aggregate models are not able to capture the more detailed information available by the individual measurements. Also, the case of a bus with no measurements has not been addressed in the previous work. Hence, the introduction of these new circuit models represents the main contribution of this paper, providing a complete, generalized modelling framework for ECF-based state estimation. For each set of RTU data, two circuit models are proposed, in order to ensure compliance with the different modelling frameworks used by the NECF-SE and LECF-SE estimators. An important but reasonable assumption is that all RTU devices measure and communicate voltage magnitude, as voltage measurements are necessary to calculate different parameters that are needed for the implementation of the ECF-based estimators. All proposed models are applicable whether the set of RTU data comprises active and reactive power measurements or current magnitude and power factor data.
\vspace{-0.2cm}
\subsection{Line Flow Measurements in One Branch} \label{sec: 4a}
In case an RTU located at a bus connected to $n$ lines obtains voltage magnitude and active and reactive line flows in only one of the lines connected to the bus, this measurement set can be mapped to the sub-circuits presented in Fig. \ref{fig:1lineflow}. The nonlinear model shown in Fig. \ref{fig:1lineflow_int} is applicable if the NECF-SE method is used, while the linear circuit model from Fig. \ref{fig:1lineflow_lin} is compliant with the LECF-SE estimator. 

In order to leverage the available set of measurements and the modelling approaches presented in Figs. \ref{fig:RTU_interval_inj}-\ref{fig:RTU_lin_model}, the RTU bus is split into 2$n$ nodes within the circuit framework, $n$ nodes each in the real and the imaginary sub-circuit. Within each sub-circuit, real or imaginary, each of these nodes is connected to one of the $n$ lines, i.e. connected to the circuit that models the corresponding line and its connection to the rest of the grid. Line flow measurements are then transformed to the appropriate injection sub-circuits, same as those in Figs. \ref{fig:RTU_interval_inj}-\ref{fig:RTU_lin_model}, because these measurements provide the information about the apparent power that is either injected into or withdrawn from the line. For the lines without measurements no actual measurement model is connected. However, the model needs to incorporate that all $n$ nodes in the real sub-circuit are connected and therefore need to have the same voltage $V_R$. For this reason, controlled voltage sources equal to the voltage at the node connected to the real sub-circuit of the monitored line are added to all other nodes within the real sub-circuit. The same reasoning is used in the imaginary sub-circuit.
\begin{figure*}[t!]
    \centering
    \begin{subfigure}[h]{0.2\textwidth}
        \centering  
        \includegraphics[width=\textwidth]{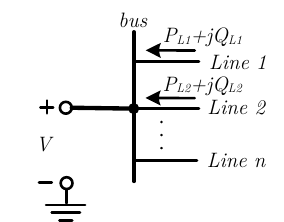}
        \caption{\hspace*{0.25em}}  
    \end{subfigure}
    \hfill
    \begin{subfigure}[h]{0.75\textwidth}  
        \centering 
        \includegraphics[width=\textwidth]{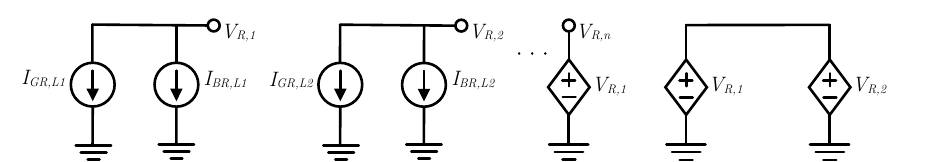}
        \caption{\hspace*{0.25em}}  
        \label{fig:2lineflow_lin_real}
    \end{subfigure}
    \caption{Circuit model for the case of line flow measurements in two branches: (a) measurement data; (b) linear real sub-circuit.}
    \label{fig:2lineflow}
    \vspace{-0.3cm}
\end{figure*}

The nonlinear model in Fig. \ref{fig:1lineflow_int} is applicable if the NECF-SE estimator is used. In accordance with this method, conductance $G_L$ and susceptance $B_L$ are bounded variables, and their upper and lower limits are obtained by using interval arithmetic and the following relations:
\begin{align}
    G_{L} &= \frac{P_L}{V^2} = \frac{I_{L}}{V}\text{cos}\left(\phi_{L}\right)\label{eq:G_RTU}\\ 
    B_{L} &= \frac{Q_L}{V^2} = \frac{I_{L}}{V}\text{sin}\left(\phi_{L}\right)\label{eq:B_RTU}
\end{align}
where $V$ is the measured bus voltage magnitude, $P_L$ and $Q_L$ are measured active and reactive line flows, $I_L$ is the measured line current magnitude and $\phi_L$ is the measured angle between bus voltage and line current. Thus, the proposed modelling approach can be used whether the set of communicated measurements consists of active and reactive branch flows, or line current magnitude and power factor data. In both cases, voltage magnitude data is necessary. Appropriate terms are added to the objective function of the nonlinear estimator for each conductive and susceptive component in the proposed circuit model, in the same way it was done in \eqref{eq: optimization} for the injection model from Fig. \ref{fig:RTU_interval_inj}. Relations \eqref{eq:G_RTU}-\eqref{eq:B_RTU} are derived for the case of load conditions. Therefore, positive values of $G_L$ indicate active power being withdrawn from the line, while the active power injection into the line renders $G_L$ negative. The same relation exists between $B_L$ and reactive line power. 

In Fig. \ref{fig:1lineflow_lin} the model for the LECF-SE method is shown. All current source values $I_{GR,L}$, $I_{GI,L}$, $I_{BR,L}$ and $I_{BI,L}$ are variables that are obtained by minimizing the weighted squares of the differences between these variables and the appropriate linear measurement functions $h(x)$, in the same way it was done in \eqref{eq: optimization2} for the injection model from Fig. \ref{fig:RTU_lin_model}. The corresponding current source-measurement function pairs are shown in Table \ref{tab:RTU_pair}. Expressions for these functions are obtained from \eqref{eq: I_R}-\eqref{eq: I_I}. Voltages $V_R$ and $V_I$ are real and imaginary voltages of the observed RTU bus, and the proposed model again ensures that $V_R=V_{R,k}$ and $V_I=V_{I,k}$, with $k\in\{1,...,n\}$, by adding controlled voltage sources. The proposed methodology can be implemented whether the set of measurements is based on power or current data. The given relations for $h(x)$ correspond to load conditions. Hence, positive $I_{GR,L}$ and $I_{GI,L}$ indicate active power withdrawal from the line, while their negative values imply active power injection into the line. The same analogy stands for $I_{BR,L}$ and $I_{BI,L}$ and reactive line power.

\subsection{Line Flow Measurements in more than One Branch}
The modelling methodology presented in Sect. \ref{sec: 4a} can be further extended in case line flow measurements are available in more than one line connected to a bus equipped with an RTU device. The appropriate circuit model for a bus connected to $n$ lines is shown in Fig. \ref{fig:2lineflow}. Due to space limitations, only the linear real sub-circuit is presented, but the same approach is used to derive the imaginary sub-circuit, as well as the nonlinear circuit model. 
\begin{table}[t!]
\centering
\caption{RTU Linear Measurement Functions}
\label{tab:RTU_pair}
\scalebox{1.2}{
\begin{tabular}{ |c|c| }
 \hline
 \textbf{Variables} & $\boldsymbol{h(x)}$\\ 
 \hline
 $I_{GR,L}$ & $(P_L/V^2)V_R$ \text{or} $(I_L/V)\text{cos}\left(\phi_L\right)V_R$\\ 
 \hline
 $I_{BR,L}$ & $(Q_L/V^2)V_I$ \text{or} $(I_L/V)\text{sin}\left(\phi_L\right)V_I$\\ 
 \hline
 $I_{GI,L}$ & $(P_L/V^2)V_I$ \text{or} $(I_L/V)\text{cos}\left(\phi_L\right)V_I$\\ 
 \hline
 $I_{BI,L}$ & $(Q_L/V^2)V_R$ \text{or} $(I_L/V)\text{sin}\left(\phi_L\right)V_R$\\ 
 \hline
\end{tabular}}
\vspace{-1em}
\end{table}
\begin{figure*}
\vspace{-0.5cm}
    \centering
    \begin{subfigure}[h]{0.25\textwidth}
        \centering  
        \includegraphics[width=\textwidth]{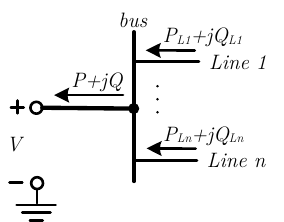}
        \caption{\hspace*{0.2em}}  
    \end{subfigure}
    \hfill
    \begin{subfigure}[h]{0.35\textwidth}  
        \centering 
        \includegraphics[width=\textwidth]{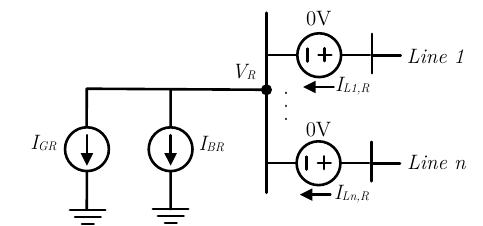}
        \caption{\hspace*{0.25em}}  
        \label{fig:3lineflow_inj_real}
    \end{subfigure}
        \hfill
    \begin{subfigure}[h]{0.35\textwidth}  
        \centering 
        \includegraphics[width=\textwidth]{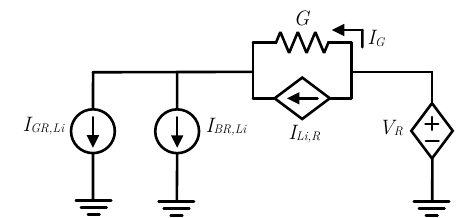}
        \caption{\hspace*{0.25em}}  
        \label{fig:3lineflow_line_real}
    \end{subfigure}
    \caption{Circuit model for an RTU bus with injection and line flow measurements: (a) measurement data; (b) linear real injection sub-circuit; (c) linear real line flow sub-circuit.}
    \label{fig:3lineflow}
    \vspace{-0.3cm}
\end{figure*}

Let $\Omega_L$ and $\Omega_{NL}$ be the sets of nodes in the real sub-circuit that correspond to the lines with and without line flow measurements, respectively. Furthermore, let $N_{\Omega_L}$ be the number of nodes in $\Omega_L$. As before, the line flow measurements obtained at each line, in combination with the voltage magnitude measurement, are represented by an injection model connected to the respective node from the set $\Omega_L$. Controlled voltage sources are connected to the nodes in the set $\Omega_{NL}$, with their values equal to the voltage of any node from the set $\Omega_L$. The main novelty, compared to the model from \ref{sec: 4a}, is the addition of another circuit that is separated from the rest of the system's equivalent circuit. In this circuit, $N_{\Omega_L}$ controlled voltage sources are connected in parallel. The value of each of these sources is controlled by a different node voltage from the set $\Omega_L$. In this way, it is ensured that all $n$ nodes within the real sub-circuit have equal voltages, thus accurately representing the real system. All explanations regarding the handling of circuit variables and their inclusion in the respective objective functions of the NECF-SE and LECF-SE estimators are the same as in \ref{sec: 4a}.

The proposed methodology still applies if line flows are being monitored in all lines connected to an RTU bus, along with the bus voltage magnitude. In that case, the set $\Omega_{NL}$ is obviously empty, as $N_{\Omega_L}=n$. The available set of measurements is modelled with $n$ injection sub-circuits that are connected to their respective transmission line sub-circuits. In order to ensure that all $n$ nodes have equal voltages, $n$ dependent voltage sources are connected in parallel at an additional node. The value of each of these sources is equal to a different node voltage from the set $\Omega_L$.    

\subsection{RTU bus with Injection and Line Flow Measurements}
Another possible case in practice is an RTU bus equipped with both injection and line flow measurements. In order to leverage all of the available information, the circuit model shown in Fig. \ref{fig:3lineflow_inj_real}-\ref{fig:3lineflow_line_real} is proposed. Again, due to space limitations, only the linear real sub-circuit is presented and explained, but the same approach is used to derive the linear imaginary sub-circuit, as well as the nonlinear circuit model.

An RTU bus connected to $n$ lines is observed. Injected active and reactive power data, along with the measurement of the bus voltage magnitude, can be represented by an injection model comprising current sources $I_{GR}$ and $I_{BR}$, as shown in Fig. \ref{fig:3lineflow_inj_real}. Let $\Omega_{line}$ be the set of all lines in which line flows are monitored. The current injected into or withdrawn from the real sub-circuit of each line $i\in\Omega_{line}$, i.e. $I_{Li,R}$, is monitored by adding an ammeter in terms of zero-volt voltage source between the real node and the sub-circuit of the respective line.

For each line $i\in\Omega_{line}$, a separate sub-circuit is added as shown in Fig. \ref{fig:3lineflow_line_real}. The purpose of this separate sub-circuit is to relate the current flowing into or from the real sub-circuit of the line, monitored by an ammeter, with the corresponding line flow measurements. The set of line flow and bus voltage magnitude measurements is represented by an injection model consisting of current sources $I_{GR,Li}$ and $I_{BR,Li}$. Values of these current sources are variables and are treated within the estimation algorithm in the same way as it was explained in Sect. \ref{sec: 4a}. In order to accurately represent the actual system, the sum of their currents must be equal to the current withdrawn from the line, and the voltage at their joint terminal must be $V_R$. In order to model this, an approach similar to the PMU modelling is used. A controlled voltage source equal to the real node voltage $V_R$ is added in series to the current sources $I_{GR,Li}$ and $I_{BR,Li}$, along with the parallel connection of the current source controlled by the value of current flowing through the respective ammeter $I_{Li,R}$ and a conductance $G$. The purpose of this conductance is to preserve the effect of both the controlled current source $I_{Li,R}$ and the controlled voltage source $V_R$, similar as in the PMU model in Fig. \ref{fig:PMU_inj}. 

Since the sum of $I_{GR,Li}$ and $I_{BR,Li}$ should be $I_{Li,R}$, and the voltage at their joint terminal must be $V_R$, the square of the current flowing through the conductance $G$, i.e. $I^2_G$, must be minimized, which is added as an additional term in the objective function of the SE algorithm. As current flowing through a conductance is proportional to its value, a large value is selected for $G$, same as for $G_{PMU}$ in the PMU circuit model \cite{Jovicic2}, to ensure that a sufficiently large weight factor is assigned to this additional term, hence ensuring that the sub-circuit in Fig. \ref{fig:3lineflow_line_real} accurately represents the real system.  

\subsection{Non-Monitored Buses and Null Injections}
\begin{figure}[t!]
\centering
\begin{subfigure}{0.4\textwidth}
\includegraphics[width=\textwidth]{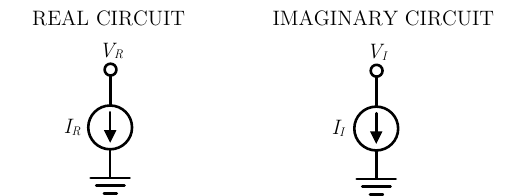}
\end{subfigure}
\caption{Circuit model for a bus with no measurements.}
\label{fig:no_meas}
\vspace{-0.2cm}
\end{figure}
A bus that is not equipped with any kind of measuring device can be represented within the circuit framework by a circuit model presented in Fig. \ref{fig:no_meas}. Current sources are added to the corresponding nodes within the real and imaginary sub-circuits, thus modelling the active and reactive power generation or consumption at this bus. Both node voltages $(V_R, V_I)$ and currents of the current sources $(I_R, I_I)$ are variables that are determined by the estimation algorithm. Since there is no available information about the bus, this circuit model does not affect the estimator's objective function.

Finally, a bus with null injections is modelled by omitting the current sources shown in Fig. \ref{fig:no_meas}, since neither generation nor consumption exist. Therefore, the corresponding nodes in the real and imaginary sub-circuits are only connected to the sub-circuits of the lines incident to the respective bus. 
            
\section{Numerical Results} \label{sec: 5}
Performance of the following three state estimators will be evaluated and compared in this section:
\begin{itemize}
    \item NECF-SE \cite{Jovicic};
    \item LECF-SE \cite{Jovicic2};
    \item Hybrid Constrained State Estimator (HC-SE), based on the conventional WLS approach \cite{Valverde2011}.
\end{itemize}

Both ECF-based estimators are considered with the extended modelling frameworks, provided by the circuit models proposed in this paper. Since both of these algorithms are essentially constrained estimators, the HC-SE algorithm that is based on the conventional WLS approach is selected for comparison, as one of the well-known recently proposed methods that is also constrained and handles PMU and RTU measurements simultaneously. The performance of the algorithms is compared by solving several test cases of different sizes. The influence of high PMU penetration levels is also examined.

\subsection{Performance Comparison for Small Systems} \label{subsec: small}
Performance of the algorithms is assessed using the IEEE 14, 57 and 118 bus test cases. For each test system, the available measurement set comprises both PMU and RTU measurements. The set of RTU measurements consists of voltage magnitude and active and reactive power injection and power flow data. The PMU measurements are assumed to comprise voltage and current magnitudes and angles, which is a common case in practice. However, PMU data in polar form do not comply with ECF-based estimators that are based on current and voltage variables in rectangular form. Therefore, in order to incorporate PMU data into ECF-based estimators, each PMU magnitude ($Z$) and angle ($\theta_z$) measurement must be transformed to a rectangular form, given by $z_{real}$ and $z_{imag}$, where the variances of the resulting rectangular measurements can be calculated based on the following relations \cite{Bi2008}: 
\begin{align}
    \sigma^2_{z_{real}} &= \left(\text{cos} \theta_z\right)^2\sigma^2_{Z}+\left(Z\cdot\text{sin} \theta_z\right)^2\sigma^2_{\theta_z}\\
    \sigma^2_{z_{imag}} &= \left(\text{sin} \theta_z\right)^2\sigma^2_{Z}+\left(Z\cdot\text{cos} \theta_z\right)^2\sigma^2_{\theta_z}
\end{align}
where $\sigma^2_Z$ and $\sigma^2_{\theta_z}$ are variances of $Z$ and $\theta_z$, respectively.

The locations of PMU buses in each test case are selected to match the optimal positioning that has been determined in \cite{Valverde2011}. The measurement set composition for all test cases is summarized in Table \ref{tab:measures}. The slack bus in each system is equipped with a PMU device in order to provide a reference for the angles. Furthermore, it is assumed that each PMU has a sufficient number of channels to monitor each line incident to its bus. Full system observability is ensured for all test cases. 

\begin{table}[t!]
\centering
\caption{Measurement Sets}
\label{tab:measures}
\begin{tabular}{ |c|c|c|c|c| }
 \hline
 \multicolumn{1}{|c|}{\multirow{2}{*}{\textbf{Test Case}}} & \multicolumn{1}{c|}{\multirow{2}{*}{\textbf{\# PMU buses}}} & \multicolumn{3}{c|}{\textbf{\# RTU Measurements}}\\
 \cline{3-5}
  & &\textbf{Voltage}  & \textbf{Injection} & \textbf{Flows}\\ 
 \hline
 14 buses & 3 & 10 & 12 & 22\\
 \hline
 57 buses & 7 & 48 & 50 & 64\\ 
 \hline
 118 buses & 10 & 106 & 102 & 124\\ 
 \hline
\end{tabular}
\end{table}
All measurement errors are assumed to follow Gaussian distribution with zero mean. The selected values for standard deviations of different types of measured data are given in Table \ref{tab:deviations}. Each measurement $z$ is generated by randomly selecting a value in the range  $\left[z^t-\sigma_z, z^t+\sigma_z\right]$, where $z^t$ is the true value of the corresponding measurement, obtained from the power flow solution of the respective test system from MATPOWER, and $\sigma_z$ is the standard deviation of the measurement $z$. 

\begin{table}[t!]
\centering
\caption{Measurement Standard Deviations}
\label{tab:deviations}
\begin{tabular}{|c|c|c|c|c|c|}
\hline
\multicolumn{3}{|c|}{\textbf{RTU Measurements}} & \multicolumn{3}{c|}{\textbf{PMU Measurements}}\\
\hline
\textbf{Voltage}   & \textbf{Injection}   & \textbf{Power Flows}  & \textbf{Voltage}   & \textbf{Current}   & \textbf{Angle}\\
 \hline
0.2\%   &  1\% & 1\%   & 0.02\%  & 0.02\%  & $0.01^o$\\
\hline
\end{tabular}
\vspace{-0.3cm}
\end{table}
Two different performance indices are used to quantify the accuracy of the evaluated estimation algorithms. The first index is the sum of the squared differences between the estimated and true states:
\begin{equation}
    \sigma^2_{x} = \sum^{2N}_{i = 1}(\hat{x}_i - x^t_i)^2
\end{equation}
where $\hat{x}$ and $x^t$ are estimated and true states, respectively, and $N$ is the number of system buses. The second index is the ratio between the variance of estimated measurements and the variance of their original values:
\begin{equation}
    \xi = \frac{\sum^{M}_{i = 1}(\hat{z}_i - z^t_i)^2}{\sum^{M}_{i = 1}(z^m_i - z^t_i)^2} \label{eq: meas_err}
\end{equation}
where $\hat{z}$, $z^m$ and $z^t$ are estimated, original and true measurement values, respectively, and $M$ is the number of measurements.  

All three estimators are implemented in Matlab, and simulations are carried out on a PC with an Intel Core i7-6600 CPU and 16 GB of RAM. The SNOPT solver is used in order to solve the optimization problem associated with NECF-SE \cite{SNOPT}. For each test case and estimation method, a total of 100 simulations are performed for different values of measurement errors, and the average values of the performance indices are given in Table \ref{tab:results}. The numerical results show that all three estimators have a very similar level of accuracy. The NECF-SE estimator performs the best for 14 and 57 bus test systems. However, it is slightly less accurate than the other two methods for the 118 bus test system. The HC-SE estimator exhibits a rather small advantage in terms of accuracy over LECF-SE for all test systems. It should be noted that the measured PMU data are given in polar form and thus have to be transformed to rectangular coordinates in order to be used with ECF-based estimators, which slightly deteriorates their performance. Based on the obtained values of the index $\xi$ for all three methods, it is obvious that the variance of the estimated measurements is significantly smaller than the variance of the original measured data.

\begin{table}[t]
\centering
\caption{Average Performance Indices}
\label{tab:results}
\begin{tabular}{ |c|c|c|c| }
 \hline
 \textbf{Test Case} & \textbf{Method} & $\boldsymbol{\sigma^2_x}$ & $\boldsymbol{\xi}$\\ 
 \thickhline
 \multicolumn{1}{|c|}{\multirow{3}{*}{14 buses}} & NECF-SE & 1.2515\,x\,$10^{-7}$ & 0.0164\\ 
 \cline{2-4}
 & LECF-SE & 4.7422\,x\,$10^{-7}$ & 0.1040\\ 
 \cline{2-4}
 & HC-SE & 3.4187\,x\,$10^{-7}$ & 0.0982\\ 
 \thickhline
  \multicolumn{1}{|c|}{\multirow{3}{*}{57 buses}} & NECF-SE & 5.6419\,x\,$10^{-6}$ & 0.1073\\ 
 \cline{2-4}
 & LECF-SE & 9.8294\,x\,$10^{-6}$ & 0.1877\\ 
 \cline{2-4}
 & HC-SE & 7.6112\,x\,$10^{-6}$ & 0.1418\\ 
 \thickhline
  \multicolumn{1}{|c|}{\multirow{3}{*}{118 buses}} & NECF-SE & 5.2834\,x\,$10^{-5}$ & 0.3335\\ 
 \cline{2-4}
 & LECF-SE & 2.9356\,x\,$10^{-5}$ & 0.2661\\ 
 \cline{2-4}
 & HC-SE & 1.4210\,x\,$10^{-5}$ & 0.1839\\ 
 \hline
\end{tabular}
\vspace{-0.2cm}
\end{table}
The average computational time for all three methods is summarized in Table \ref{tab:time}. LECF-SE has a substantial computational advantage compared to the other two methods. This stems from the fact that it estimates the state of the system by solving a set of linear equations. Poor computational speed of the NECF-SE estimator is mainly a consequence of the use of commerical nonlinear solver. However, it can be significantly improved by utilizing a circuit approach to solve the NECF-SE's optimization problem instead \cite{Jereminov}.
\begin{table}[t]
\centering
\caption{Average Computational Time}
\label{tab:time}
\begin{tabular}{ |c|c|c|c| }
 \cline{2-4}
 \multicolumn{1}{c|}{}& \textbf{14 buses} & \textbf{57 buses} & \textbf{118 buses}\\ 
 \hline
 \textbf{NECF-SE} & 0.4667s & 2.1435s & 8.4393s\\ 
 \hline
 \textbf{LECF-SE} & 0.0001s & 0.0004s & 0.0008s\\ 
 \hline
 \textbf{HC-SE} & 0.0157s & 0.0739s & 0.2304s\\ 
 \hline
\end{tabular}
\end{table}
\subsection{Performance Comparison for Large Systems}
To evaluate the scalability of the algorithms, the 2869 and 13659 bus systems provided by the PEGASE project are used \cite{13k_buses}. The measurement sets are generated in the same way as in \ref{subsec: small}, with values of standard deviations of different types of measurements selected according to Table \ref{tab:deviations} and 6\% of the buses being covered by PMU measurements. Full system observability is ensured. 

The obtained results again demonstrate a very similar accuracy of LECF-SE and HC-SE algorithms. However, the difference in their computational times becomes significant, as the HC-SE requires more than 6 seconds to converge for the case of 13659 bus test system, while the LECF-SE provides results in less than 1 second. The clear advantage of LECF-SE in terms of scalability is therefore verified. The NECF-SE solved by nonlinear commercial solver proved to not be competitive with respect to computational speed. As an alternative, solving its optimization problem within the equivalent circuit framework is suggested to reduce the computational complexity, achieving significantly improved computational speed \cite{Jereminov}. For consistency and given the fact that it is still slower than the LECF-SE, we omit the discussion of using this alternative solver.

\subsection{Full Observability by PMU Measurements}
Another group of simulations is performed to apply LECF-SE and HC-SE methods to 14, 57 and 118 bus test systems with 100\% PMU penetration, i.e. with all buses being equipped with a PMU device. Again, the computational time is averaged over 100 simulations and the average values are shown in Table \ref{tab:timePMU}. Comparison of these results and the values obtained in Table \ref{tab:time} for the measurement allocation described in Table \ref{tab:measures} demonstrates that the HC-SE method converges slower as the number of PMU buses increases, while the computational speed of LECF-SE remains unaffected. The performance of the HC-SE method is degraded mainly because of current phasor measurements, as the increase of their number results in an increased dimension of the gain matrix and its ill-conditioning.   

\begin{table}[t]
\vspace{-0.1cm}
\centering
\caption{Average Computational Time for 100\% PMU Penetration}
\label{tab:timePMU}
\begin{tabular}{ |c|c|c|c| }
 \cline{2-4}
 \multicolumn{1}{c|}{}& \textbf{14 buses} & \textbf{57 buses} & \textbf{118 buses}\\ 
 \hline
 \textbf{LECF-SE} & 0.0001s & 0.0004s & 0.0008s\\ 
 \hline
 \textbf{HC-SE} & 0.0673s & 0.3994s & 2.7253s\\ 
 \hline
\end{tabular}
\vspace{-0.5cm}
\end{table}
\vspace{-0.2cm}
\section{Conclusion} \label{sec: 6}
This paper introduces new circuit models to provide a comprehensive modelling framework for the previously proposed ECF-based state estimators. New models are derived for different combinations of conventional measurements. Also, the circuit representation of null injections and buses with no measurements is presented. The performance comparison of the two ECF-based estimators with enhanced modelling frameworks and a well-known hybrid constrained estimator, performed by running simulations on a range of test systems, demonstrates that all three methods have a comparable level of accuracy. However, the numerical results exhibit a significant advantage of the LECF-SE method in terms of computational speed, enabled by its fully linear circuit representation of the system along with measurements. This is particularly true for increasing levels of PMU penetration. 

\bibliographystyle{IEEEtran}

\end{document}